**Magnetic coupling of porphyrin molecules through graphene**

By *Christian F. Hermanns, Kartick Tarafder, Matthias Bernien, Alex Krüger, Yin-Ming Chang, Peter M. Oppeneer*, and *Wolfgang Kuch\**

[*]     C. F. Hermanns, Dr. M. Bernien, A. Krüger, Dr. Y.- M. Chang, Prof. Dr. W. Kuch

Institut für Experimentalphysik

Freie Universität Berlin

Arnimallee 14, 14195 Berlin, Germany

E-mail: kuch@physik.fu-berlin.de

Dr. K. Tarafder, Prof. P. M. Oppeneer

Department of Physics and Astronomy

Uppsala University

P. O. Box 516, 75120 Uppsala, Sweden

Graphene, a material with unique properties,[1] is expected to complement todays Si-based information technology with new and more efficient functions.[2-4] It exhibits desirable properties for spin electronic applications such as high charge carrier mobility, low intrinsic spin-orbit interaction, as well as low hyperfine interaction.[5-6] In particular, magnetic molecules in contact with graphene constitute a tantalizing approach towards organic spin electronics because of the reduced conductivity mismatch at the interface. In such a system a bit is represented by a single molecular magnetic moment, which must be stabilized against thermal fluctuations.[7] Here, we show in a combined experimental and theoretical study that the moments of paramagnetic Co-octaethylporphyrin (CoOEP) molecules on graphene can be aligned by a remarkable antiferromagnetic coupling to a Ni substrate underneath the graphene. This coupling is mediated via the π electronic system of graphene, while no covalent bonds between the molecule and the substrate are established, and the molecules sit at a distance of ~3.3 Å above the graphene plane. The laterally extended π electron system of graphene exhibits metal-like electronic properties along the plane, and molecule-like properties perpendicular to the plane, which makes graphene highly relevant for the design of hybrid metal-organic spintronic materials. Several studies have focused on the spin-split electronic states of graphene or the injection of spin-polarized currents from metallic ferromagnetic electrodes.[8-10] Electronic spin transport and spin precession have been observed in graphene over micrometer distances even at room temperature.[11, 12] Molecules adsorb on graphene mainly by van der Waals interaction, thereby the molecular properties of the adsorbate, like its reactivity, are preserved. This opens the door to an optimization of the undisturbed molecular design of an adsorbate independently of its interaction with the substrate.

We use X-ray magnetic circular dichroism (XMCD) in the absorption of soft X rays as an element-selective probe of the magnetization of CoOEP molecules in the submonolayer regime on a graphene-passivated Ni film,[13] supported by a W(110) single crystal surface. In the upper and lower panel of **Fig. 1a**, we show Co $L_{2,3}$ X-ray absorption (XA) spectra and XMCD spectra of 0.7 molecular monolayers (ML) of CoOEP on top of graphene/Ni, respectively, measured in zero field at remanence of the ferromagnetic Ni film. The astonishing presence of a fininite Co XMCD signal proves on the one hand a net $3d$ magnetic moment localized on the Co ions and on the other hand an unexpected magnetic coupling between the magnetization of the Ni layer and these Co moments, stabilizing them against thermal fluctuations. The XA spectrum as well as the XMCD spectra exhibit a particular finestructure at the Co $L_3$ edge, which, by comparison to CoPc bulk measurements,[14] is consistent with a $d^7$ low-spin state of Co. The XMCD difference curves in Fig. 1a show a positive excursion at the Co $L_3$ edge between about 777 and 781 eV photon energy, which is about a factor 4.1 larger at 30 K compared to 130 K, and only a small negative signal at the Co $L_2$ edge at around 793.5 eV. The sign of this XMCD signal is opposite to that of the ferromagnetic Ni layer underneath the graphene, displayed in the inset of the upper panel of Fig. 1a. This proves an antiferromagnetic coupling between the magnetic moments of the Co centers and the Ni magnetization.

By employing the XMCD spin sum rule on the Co XMCD spectrum measured at 30 K,[15] an effective spin magnetic moment of $(0.87 \pm 0.05)\ \mu_B$ is obtained. As the Co magnetic moments are subject to thermal fluctuations at finite temperatures, this value presents a lower limit. The effective spin magnetic moment fits to a low-spin $S = 1/2$ configuration of the Co ions. The coupling energy $E_{ex}$ between the Co spins and the Ni substrate can be estimated from the Co magnetization relative to the one of Ni as a function of temperature, assuming that all molecules interact in the same manner with the surface. In **Fig. 1b** the Co and Ni $L_3$ XMCD signals, normalized to their extrapolated saturation values, are plotted vs. temperature. The temperature progression of the Co and Ni magnetizations clearly differs. Experimental data are shown together with a theoretically modeled temperature dependence of the respective magnetization. A Brillouin function $B_J(\alpha)$ with $\alpha = E_{ex}/(k_B T)$, temperature $T$, and Boltzmann constant $k_B$ is used to represent the relative Co magnetization, incorporating the coupling to the magnetic substrate as an effective magnetic field and neglecting magnetic anisotropy:[16]

$$M_r^{Co}(T) = M_r^{Ni}(T) B_J(\alpha). \tag{1}$$

As an exchange field acts directly on the spin, $J = 1/2$ and a temperature-independent coupling energy $E_{ex}$ is assumed. The fit yields $E_{ex} = (1.8 \pm 0.5)$ meV for the data acquired at temperatures between 30 and 130 K. Thereby the Ni magnetization is modeled by a $(1-T/T_C)^\beta$ law, with a Curie temperature of $T_C = 630$ K and a critical exponent $\beta = 0.365$.[17] The strength of the magnetic coupling across the graphene layer as opposed to metalloporphyrins adsorbed directly on reactive ferromagnetic substrates is comparatively small.[16, 18]

To shed light on the unexpected exchange coupling of paramagnetic metalorganic molecules mediated through graphene, we have performed *ab initio* calculations employing the DFT+$U$ approach (see Methods section for details). In recent studies of graphene on Ni(111) six possible arrangements of the graphene atoms on a Ni(111) surface have been discussed, the most prominent ones are denoted as the bridge-top, top-fcc, top-hcp, and fcc-hcp configurations.[19-23] For these four configurations, the three more symmetric ones and the bridge-top configuration that had been favored by DFT calculations,[20, 24] we have performed geometrical optimizations and *ab initio* calculations of the molecule-substrate exchange interaction. Our calculations were performed both with and without Van der Waals (VdW) correction terms.[25] The bridge-top configuration has the lowest computed total energy;[20] consistently, we find that this adsorption geometry gives agreement between measured and calculated results. Also for hcp-fcc adsorption geometry an antiparallel alignment of Ni and Co spins is computed, however, this arrangement leads to a higher total energy. Therefore, only the results computed for the bridge-top geometry are analyzed further in the following. For completeness' sake we present the computed electronic and magnetic properties of the system also for the three other graphene-Ni adsorption geometries in the Supporting Information (see Fig. S6).

The porphyrin molecule is computed to adsorb in planar geometry on graphene, with a slight bending of the macrocyclic rings towards the substrate. This is consistent with results of angle-dependent XA measurements at the carbon and nitrogen $K$ edge (see Fig. S3 and S4 in the Supporting Information). The VdW correction has a small influence on the equilibrium distances between the Ni top layer and graphene, and between graphene and the central metal ion. The Ni-graphene distance is 2.08 Å, whereas the graphene-metal ion distance is 3.68 Å (without VdW) and 3.51 Å (with VdW) for the bridge-top configuration. The VdW corrections thus reduce the molecule-graphene distance slightly (~5%), while interatomic distances in the porphyrin molecule are not affected. The computed optimal planar position of the Co ion is near the center of the hexagon formed by graphene C atoms.

Our *ab initio* DFT+$U$ calculations predict, in the equilibrium adsorption position, a low spin $S = 1/2$ state of the Co ion and an antiparallel alignment of its magnetic moment with respect to the Ni magnetization, consistent with experiment. **Fig. 2** shows an atom-projected and spin-resolved density-of-states-plot for the Co porphyrin/graphene/Ni system, which additionally reveals a small spin polarization on the N atoms, antiparallel to that of the Co ions. Unraveling further the origin of the coupling behavior, we consider the *d*-orbital occupations of the central ion and the resulting magnetization density. The seven 3$d$ electrons of the Co ion are distributed over the three filled $d_{yz,xz}$ and $d_{xy}$ orbitals and the half-filled $d_{3z^2-r^2}$ orbital, which determines the magnetization density distribution on the Co ion (see Fig. S8 in the Supporting Information for the DOS of the Co 3$d$ orbitals). **Fig. 3a** shows the *ab initio* computed magnetization densities (with VdW) of a Co porphyrin molecule on graphene supported on Ni. The negative magnetization density on the Co center, shown in light blue, has the expected $d_{3z^2-r^2}$-type shape. At first sight one might suppose that a direct exchange coupling of Co occurs through hybridization of the $d_{3z^2-r^2}$ and graphene $p_z$ orbitals, however, a closer inspection showed that this is not the case. To shine light on the electronic wavefunction overlap in space we present in **Fig. 3b** a charge density cross-section plot. This plot illustrates that the electronic interaction between the porphyrin molecules and graphene occurs due to little overlap of macrocyclic $\pi$ and graphene $p_z$ orbitals; the overlap of the short-ranged metal 3$d$ and graphene $p$ orbitals is negligible, as can be seen from the dip in the charge density at the Co site. This is also evident from isotropic XAS measurements at the cobalt $L_3$ edge of CoOEP molecules in a polycrystalline bulk sample, on graphene/Ni/W(110), and on bare Ni films (see Fig. S5 in the Supporting Information). Considering the possible exchange paths from the Ni top layer to the metal center, we observe the following: A small negative spin density on graphene (-0.001 $\mu_B$), shown in light blue in Fig. 3a, which is antiparallel to the dominant one on Ni, is induced in the graphene upper π-bond lobe of the C – C bridge (above a surface Ni atom). The lower π-bond lobe has acquired a positive magnetization density, which interconnects to a network. This results from hybridization with spin-minority Ni *sp* states (seen as extended light blue framework) with graphene $p_z$ orbitals. A weak antiparallel coupling between graphene π and porphyrin π orbitals induces a small positive spin density, residing mainly on the pyrrolic nitrogen atoms (+0.015 $\mu_B$), being thus parallel to the 3d spin density on Ni. This small parallel spin-density on the nitrogen atoms can be recognized in Fig. 3a by the yellow contours. The final chain in the exchange path is the magnetic coupling between the nitrogen atoms and the central Co ion. The N *p* orbitals hybridize weakly with the Co $d_{3z^2-r^2}$ orbital favoring an antiparallel spin polarization on N and Co. Consequently, our *ab initio* calculations unveil an indirect-direct double exchange interaction between the top-layer Ni spin and the central Co ion's spin: The graphene π-bonded sheet mediates a weak superexchange between the spin polarization of Ni and the pyrrolic nitrogens; the spin densities of the latter couple through direct exchange to the spins on the central ion of the molecule.

Magnetism of a Cr-containing molecule on a graphene sheet has been predicted, however, without a stabilizing mechanism for the spin, rendering the system to be paramagnetic.[7] Also, for graphene in contact with pure metal layers magnetism was predicted,[26] but an experimental confirmation is missing. Here, the graphene layer, on the one hand, decouples the molecules from the substrate and passivates the Ni surface.[13] Similar organic molecules as the here-studied ones, Fe phthalocyanines, have been also found decoupled electronically from the substrate on a graphene-covered metal surface, where they maintained their molecular electronic properties.[27] This virtue of a weak electronic interaction permits to achieve design of molecular functionalities of an adsorbate undisturbed by its interaction with the substrate. On the other hand, the graphene layer also mediates a magnetic interaction between the molecules and the substrate, an essential ingredient for the use of paramagnetic molecules as building blocks of a molecular spin electronics. Such molecules had moved into the center of interest after it was shown that the spin of adsorbed metalloporphyrin molecules and Tb phthalocyanine double-deckers can be stabilized against thermal fluctuations by magnetic coupling to a ferromagnetic substrate at elevated temperatures.[16,18,28,29] Density functional theory calculations revealed that the magnetic coupling of the porphyrin molecules is of superexchange type.[16,18,30] In these cases, however, the exchange coupling was established by covalent bonds between the molecules and surface atoms forming a hybrid metal-organic interface, while in the case of graphene no covalent bond is formed. Our result encourages the pursuit of spin-electronic devices such as spin qubits or spin field-effect transistors by assembling planar paramagnetic molecules wired by graphene ribbons on a surface. Electronic transport through the molecules or switching their magnetic properties, for example, could be accomplished by taking advantage of the empty sixth coordination place.

*Experimental and theoretical details*

The graphene layer has been prepared on a Ni film deposited on a W(110) single crystal surface under ultrahigh vacuum conditions ($p = 2.0 \times 10^{-10}$ mbar), following the recipe described in Refs. [8,9]. A W(110) single crystal substrate was cleaned by flash heating under $6 \times 10^{-8}$ mbar oxygen to 1600 K for 15 min, followed by five flashes to 2300 K for 10 s each. The surface quality was checked by low-energy electron diffraction. (111)-oriented Ni films of around 5.1 nm thickness were prepared by electron-beam evaporation on the clean W(110) substrate held at room temperature, followed by annealing at 570 K for 10 min. A continuous and ordered graphene overlayer is prepared via cracking of propene gas ($C_3H_6$) at a partial pressure of $10^{-6}$ mbar for 10 min at a substrate temperature of 650 K, followed by annealing to 750 K for 10 min.[9] Co(II)-2,3,7,8,12,13,17,18-octaethylporphyrin molecules were purchased from Sigma Aldrich, and evaporated by sublimating molecular powder from a crucible at around 485 K onto the sample held at room temperature. Thicknesses and coverages were estimated using quartz microbalances integrated in the evaporators, and from comparison of the signal-to-background ratio at the respective XA edges to literature data.[31]

XA measurements were performed using X rays of the helical undulator beamline UE56/2-PGM1 of BESSY II in Berlin, with a circular degree of polarization of about 85%. XA spectra were acquired in total electron yield mode recording the sample drain current as a function of photon energy, while monitoring the incoming beam intensities by the total electron yield of a freshly evaporated gold grid. Furthermore, the XA spectra were normalized to the corresponding spectra measured of the clean substrate and scaled to 1 in the pre-edge energy region. The photon energy resolution was set to 300 meV. Magnetic measurements were carried out in remanence of the Ni film with a thickness of around 26 monolayers (ML) and an easy magnetization axis in the film plane.[32] Typical photon flux densities at the sample of about $10^{12}$ $s^{-1}cm^{-2}$ were used to prevent radiation damage, which can be excluded here from the comparison of spectra taken immediately after sample preparation and at later times. Calibration of the photon energy was carried out by means of XA measurements of Co films on a Cu(100) crystal setting the position of the Co $L_3$ edge to 778.5 eV. Identical Co $L_{2,3}$ XA and XMCD spectra, with respect to the line-shape and results of the sum rule analysis, were obtained during different beamtimes for the system under study.

*Ab initio* calculations were performed with the density-functional theory +*U* (DFT+*U*) approach, wherein the strong electron-electron Coulomb interactions that exist in the open 3*d* shell of the 3*d* ion are captured by the supplemented Hubbard and exchange constants *U* and *J*. In the present calculations *U* and *J* were taken to be 4 eV and 1 eV, respectively. These values were previously shown to provide the correct spin state for free and adsorbed metalloporphyrins.[33,34] In addition our calculations were performed both with and without Van der Waals correction terms.[25] We employed the VASP full-potential plane-wave code, in which pseudo-potentials together within the projector augmented wave method are used.[35, 36] A kinetic energy cut-off of 400 eV was employed for the plane waves. For the DFT exchange-correlation functional the generalized gradient approximation was used, in PBE parameterization.[37] The metallic surface was modeled through three atomic Ni layers (adopting the fcc Ni lattice parameter). For simplicity and to focus on the magnetic interaction in the center of the molecule, the ethylene peripheral groups were replaced by hydrogen atoms. Full geometric optimizations of the metalloporphyrin molecules, the graphene and Ni surface layer were performed, including all interatomic distances until the forces were smaller than 0.01 eV Å$^{-1}$. Recent DFT+$U$ calculations revealed that metalloporphyrins on metallic substrates may exhibit two distinct adsorption distances;[34] for metalloporphyrins on graphene we obtained only a single adsorption distance. For the bridge-top adsorption geometry of graphene on Ni(111), a geometric relaxation of the graphene/Ni system was first carried out for the top Ni layer, the graphene layer, and the Co porphine position. For the reciprocal space sampling we used 3 × 3 × 1 Monkhorst-Pack *k* points.

*Acknowledgements*

B. Zada and W. Mahler are acknowledged for their technical support, and N. Reineking and A. Bruch for help during the measurements. This work has been supported by the DFG through Sfb 658, the Swedish-Indian Research Link Programme, the C. Tryggers Foundation, and the Swedish National Infrastructure for Computing (SNIC).

[1] A. K. Geim, K. S. Novoselov, *Nat. Mater.* **2007**, *6*, 183.

[2] K. V. Emtsev, A. Bostwick, K. Horn, J. Jobst, G. L. Kellogg, L. Ley, J. L. McChesney, T. Ohta, S. A. Reshanov, J. Röhrl, E. Rotenberg, A. K. Schmid, D. Waldmann, H. B.Weber, T. Seyller *Nat. Mater.* **2009**, *8*, 203.

[3] L. Britnell, R. V. Gorbachev, R. Jalil, B. D. Belle, F. Schedin, A. Mishchenko, T. Georgiou, M. I. Katsnelson, L. Eaves, S. V. Morozov, N. M. R. Peres, J. Leist, A. K. Geim, K. S. Novoselov, L. A. Ponomarenko, *Science* **2012**, *336*, 947.

[4] H. Yang, R. V. Gorbachev, R. Jalil, B. D. Belle, F. Schedin, A. Mishchenko, T. Georgiou, M. I. Katsnelson, L. Eaves, S. V. Morozov, N. M. R. Peres, J. Leist, A. K. Geim, K. S. Novoselov, L. A. Ponomarenko, *Science* **2012**, *336*, 1140.

[5] B. Traumzettel, D. v. Bulaev, D. Loss, G. Burkard, *Nat. Phys.* **2007**, *3*, 192.

[6] D. Huertas Hernando, F. Guinea, A. Brataas, *Phys. Rev. B* **2006**, *74*, 155426.

[7] S. M. Avdoshenko, I. N. Ioffe, G. Cuniberti, L. Dunsch, A. A. Popov, *ACS Nano* **2011**, *5*, 9939.

[8] A. Varykhalov, J. Sánchez-Barriga, A. M. Shikin, C. Biswas, E. Vescovo, A. Rybkin, D. Marchenko, and O. Rader, *Phys. Rev. Lett.* **2008**, *101*, 157601.

[9] Yu. S. Dedkov, M. Fonin, U. Rüdinger, C. Laubschat, *Phys. Rev. Lett.* **2008**, *100*, 107602.

[10] W. Han, K. Pi, K. M. McCreary, Y. Li, J. J. I. Wong, A. G. Swartz, R. K. Kawakami, *Phys. Rev. Lett.* **2010**, *105*, 167202.

[11] N. Tombros, C. Jozsa, M. Popinciuc, H. T. Jonkman, B. J. Wees, *Nature* **2007**, *448*, 571.

[12] B. Dlubak, M.-B. Martin, C. Deranlot, B. Servet, S. Xavier, R. Mattana, M. Sprinkle, C. Berger, W. A. De Heer, F. Petroff, A. Anane, P. Seneor, A. Fert, *Nat. Phys.* **2012**, *8*, 557.

[13] Yu. S. Dedkov, M. Fonin, C. Laubschat, *Appl. Phys. Lett.* **2008**, *92*, 052506.

[14] S. Stepanow, P. S. Miedema, A. Mugarza, G. Ceballos, P. Moras, J. C. Cezar, C. Carbone, F. M. F. de Groot, P. Gambardella, *Phys. Rev. B* **2011**, *83*, 220401.

[15] P. Carra, B. T. Thole, M. Altarelli, X. Wang, *Phys. Rev. Lett.* **1993**, *70*, 694.

[16] M. Bernien, J. Miguel, C. Weis, Md. E. Ali, J. Kurde, B. Krumme, P. M. Panchmatia, B. Sanyal, M. Piantek, P. Srivastava, K. Baberschke, P. M. Oppeneer, O. Eriksson, W. Kuch, H. Wende, *Phys. Rev. Lett.* **2009**, *102*, 047202.

[17] Y. Li, K. Baberschke, *Phys. Rev. Lett.* **1992**, *68*, 1208.

[18] H. Wende, M. Bernien, J. Luo, C. Sorg, N. Ponpandian, J. Kurde, J. Miguel, M. Piantek, X. Xu, Ph. Eckhold, W. Kuch, K. Baberschke, P. M. Panchmatia, B. Sanyal, P. M. Oppeneer, O. Eriksson, *Nat. Mater.* **2007**, *6*, 516.

[19] Yu. S. Dedkov, M. Fonin, *New J. Phys.* **2010**, *12*, 125004.

[20] W. Zhao, S. M. Kozlov, O. Höfert, K. Gotterbarm, M. P. A. Lorenz, F. Vines, C. Papp, A. Görling, H.-P. Steinrück, *J. Phys. Chem. Lett.* **2011**, *2*, 759.

[21] U. Usachov, A. M. Dobrotvorskii, A. Varykhalov, O. Rader, W. Gudat, A. M. Shikin, V. K. Adamchuk, *Vacuum* **1995**, *46*, 1101.

[22] J. Lahiri, Y. Lin, F. Sette, P. Bozkurt, I. I. Oleynik, M. Batzill, *Nat. Nanotechnology* **2010**, *5*, 326.

[23] B. Bäcker, G. Hörz, *Phys. Rev. B* **2008**, *78*, 085403.

[24] M. Fuentes-Cabrera, M. I. Baskes, A. V. Melechko, M. L. Simpson, *Phys. Rev. B* **2008**, *77*, 035405.

[25] S. Grime, *J. Comp. Chem.* **2006**, *27*, 1787.

[26] B. Li, L. Chen, X. Pan, *Appl. Phys. Lett.* **2011**, *98*, 133111.

[27] J. Mao, H. Zhang, Y. Jiang, Y. Pan, M. Gao, W. Xiao, H.-J. Gao, *J. Am. Chem. Soc.* **2009**, *131*, 14136.

[28] A. Scheybal, T. Ramsvik, R. Bertschinger, M. Putero, F. Nolting, T.A. Jung, *Chem. Phys. Lett.* **2005**, *411*, 214.

[29] A. Lodi Rozzini, C. Krull, T. Balashov, J. J. Kavich, A. Mugarza, P. S. Miedema, P. K. Thakur, V. Sessi, S. Klyatskaya, M. Ruben, S. Stepanow, P. Gambardella, *Phys. Rev. Lett.* **2011**, *107*, 177205.

[30] J. B. Goodenough, *Magnetism and the Chemical Bond*, Wiley, New York **1963**.

[31] C. F. Hermanns, M. Bernien, A. Krüger, J. Miguel, W. Kuch, *J. Phys.: Condens. Matter* **2012**, *24*, 394008.

[32] M. Farle, A. Berghaus, Y. Li, K. Baberschke, *Phys. Rev. B* **1990**, *42*, 4873.

[33] P. M. Panchmatia, B. Sanyal, P. M. Oppeneer, *Chem. Phys.* **2008**, *343*, 47.

[34] Md. E. Ali, B. Sanyal, P. M. Oppeneer, *J. Phys. Chem. C* **2009**, *113*, 14381.

[35] P. E. Blöchl, *Phys. Rev. B* **1994**, *50*, 17953.

[36] G. Kresse, J. Furthmüller, *Phys. Rev. B* **1996**, *54*, 11169.

[37] J. P. Perdew, K. Burke, M. Ernzerhof, *Phys. Rev. Lett.* **1996**, *77*, 3865.

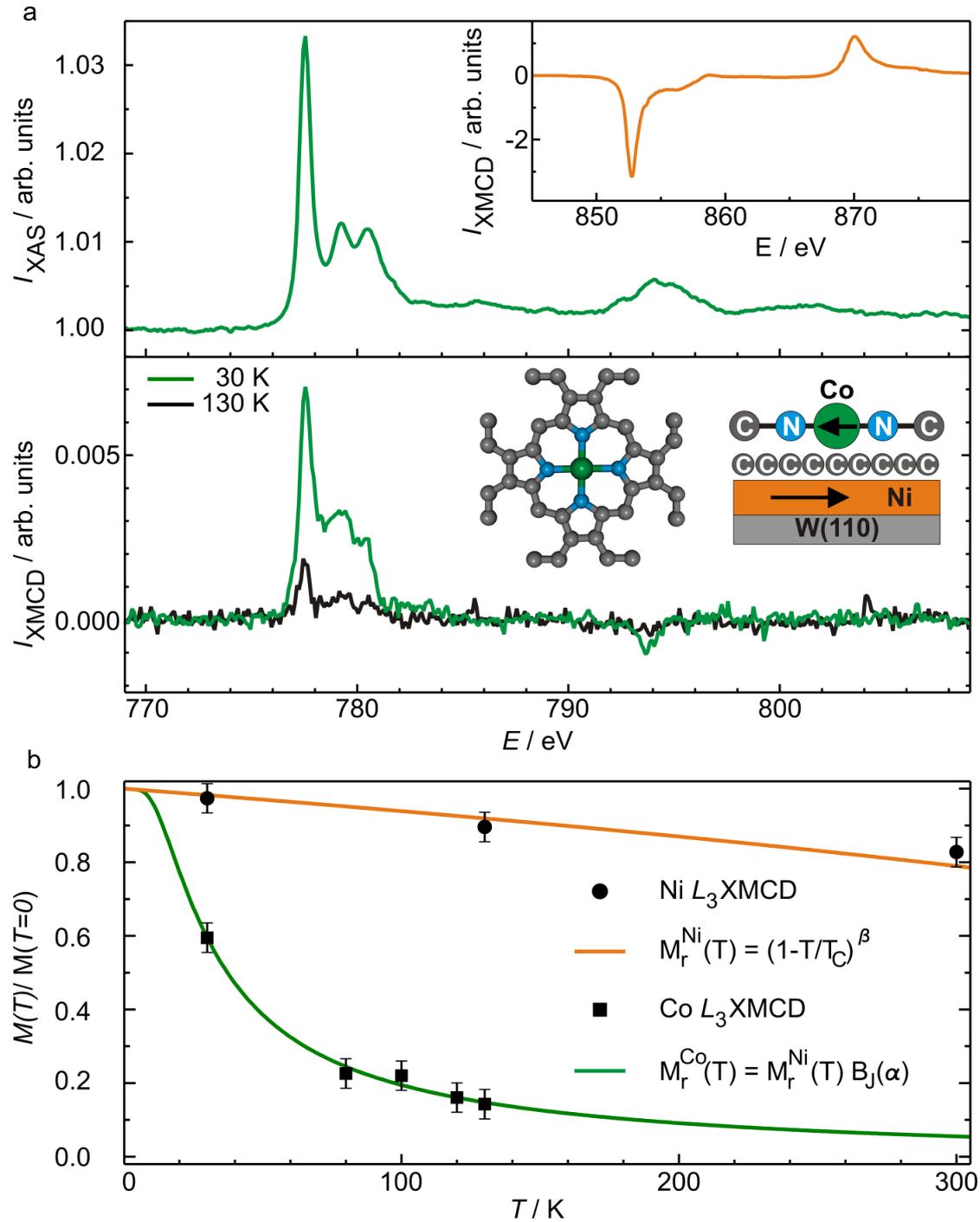

**Figure 1.** a, Co $L_{2,3}$ XAS (upper panel) and XMCD (lower panel) spectra of 0.7 ML CoOEP on graphene/Ni/W(110) measured at 70° grazing incidence at 30 K (green lines) and 130 K (black line). Insets: (upper panel) XMCD spectrum at the Ni $L_{2,3}$ edges (orange line) at 130 K showing opposite sign at the $L_3$ and the $L_2$ edges compared to the Co XMCD spectrum. (lower panel) Schematic top view of CoOEP molecule and side view of the sample, where green, blue, and grey balls represent cobalt, nitrogen, and carbon atoms, respectively, and hydrogen atoms are omitted. b, Temperature dependence of Co XMCD (squares: experimental data; green full line: fit of Brillouin-type model) and Ni XMCD (cycles: experimental data; orange full line: $(1-T/T_C)^\beta$ with $T_C$ = 630 K and $\beta$ = 0.365) for 0.7 ML CoOEP on graphene/Ni/W(110).

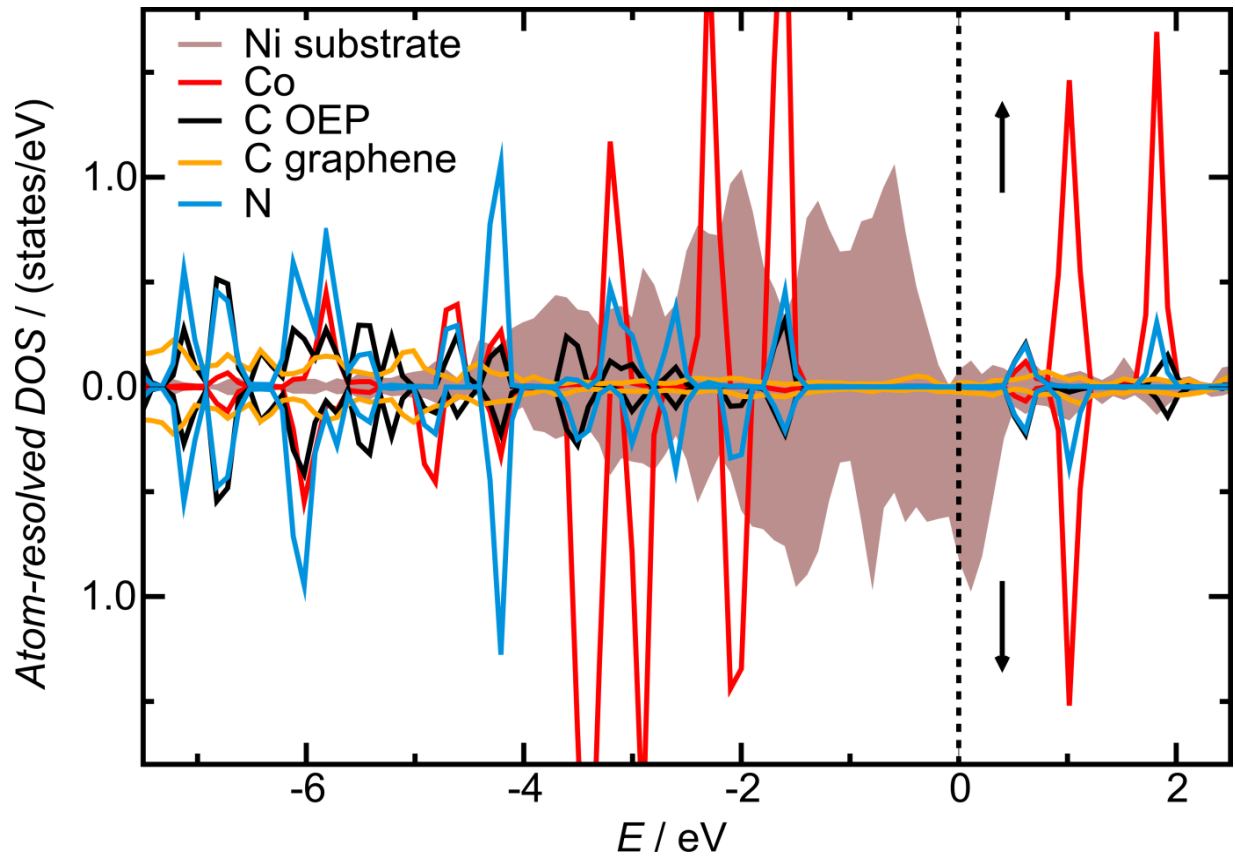

**Figure 2.** Atom-projected and spin-resolved density of states (DOS) of a Co porphyrin molecule on graphene/Ni(111) obtained from DFT+$U$ calculations. The spin polarization on the Co ions is antiparallel to that of the Ni substrate seen from the opposite shifts of the spin majority DOS. Positive atomic DOS corresponds to spin up, negative atomic DOS to spin down.

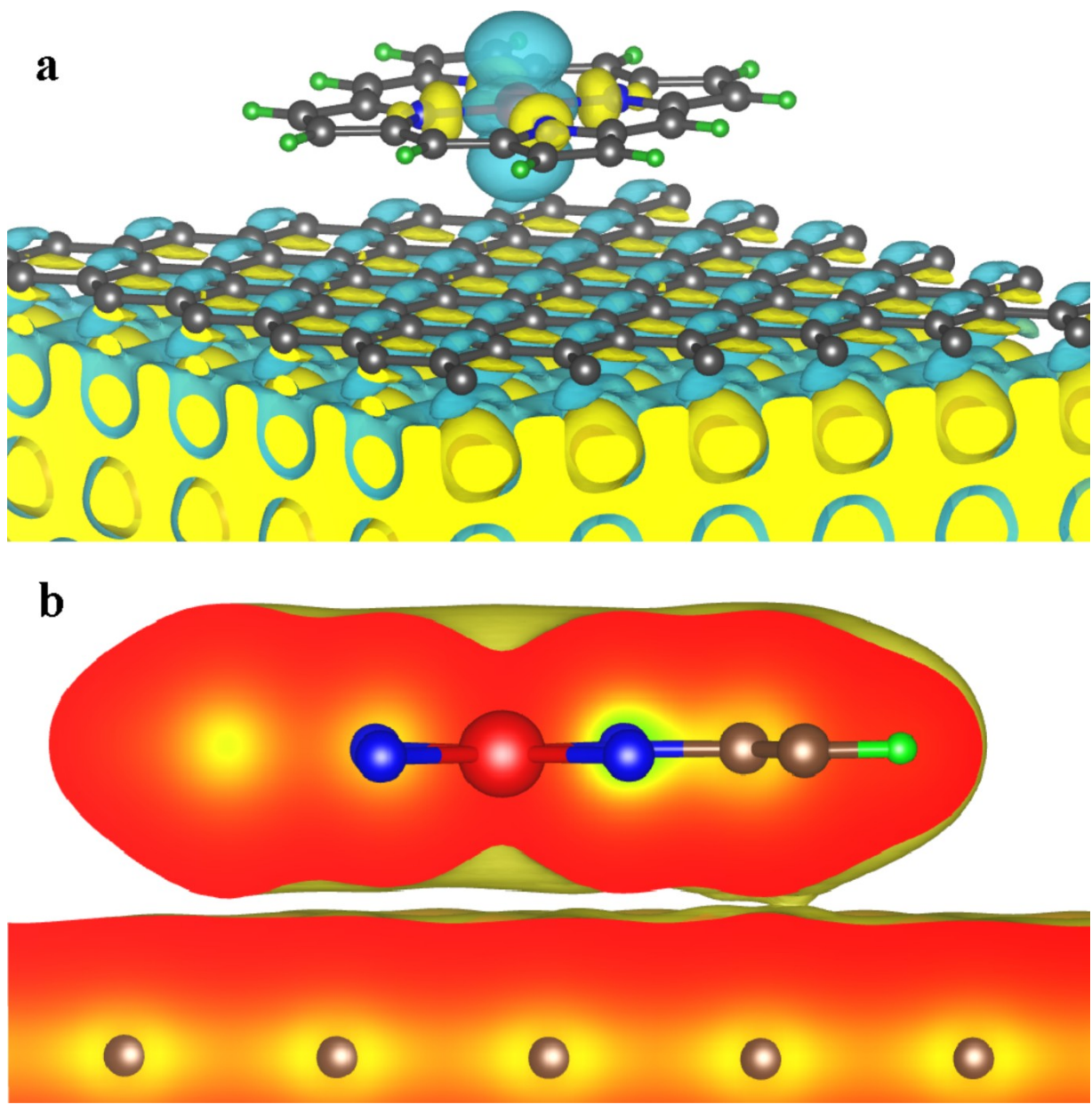

**Figure 3.** a, Calculated magnetization density of a Co porphine adsorbed on graphene/Ni. The bright yellow hypersurfaces show contours of positive magnetization densities, the light blue hypersurface shows contours of negative magnetization density. Note the small positive magnetization densities present on the nitrogen atoms. b, Charge density cross-sectional plot of a Co porphine adsorbed on graphene/Ni. The cross-section reveals a negligible overlap with the graphene charge density at the Co site. [Used isosurface values: 30 eV Å$^{-3}$]

**XMCD measurements and DFT+$U$ calculations reveal an unexpected antiferromagnetic coupling between physisorbed paramagnetic Co-porphyrin molecules and a Ni surface, separated by a graphene layer.** A positive magnetization at the Ni substrate atoms is mediated by graphene and induces a negative one at the Co site, despite of only a very small overlap between macrocyclic $\pi$ and graphene $p_z$- orbitals.

**Keywords:**

magnetic materials, X-ray absorption spectroscopy, graphene, molecular spintronics

Christian F. Hermanns, Kartick Tarafder, Matthias Bernien, Alex Krüger, Yin-Ming Chang, Peter M. Oppeneer, and Wolfgang Kuch*

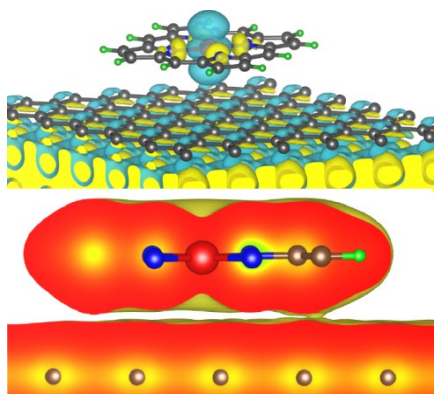

# Supporting Information

# Magnetic coupling of porphyrin molecules through graphene

By *Christian F. Hermanns, Kartick Tarafder, Matthias Bernien, Alex Krüger, Yin-Ming Chang, Peter M. Oppeneer,* and *Wolfgang Kuch\**

[*]     C. F. Hermanns, Dr. M. Bernien, A. Krüger, Dr. Y.- M. Chang, Prof. Dr. W. Kuch

Institut für Experimentalphysik

Freie Universität Berlin

Arnimallee 14, 14195 Berlin, Germany

E-mail: kuch@physik.fu-berlin.de

         Dr. K. Tarafder, Prof. Dr. P. M. Oppeneer

Department of Physics and Astronomy

Uppsala University

P. O. Box 516, 75120 Uppsala, Sweden

**Methods and sample preparation**

The additional measurements, presented in the Supplementary Information, were performed at two different beamlines at the storage ring BESSY II in Berlin: UE56/2-PGM1 and PM-3. X-ray absorption (XA) spectra were acquired in total-electron-yield mode by recording the sample drain current as a function of the photon energy. The spectra were normalized to the total electron yield of a freshly evaporated gold grid upstream to the experiment. Furthermore, the XA spectra were normalized to the corresponding spectra measured at the clean substrate and scaled to 1 in the pre-edge energy region. Data were collected with an energy resolution of 300 meV at the Co $L_{2,3}$ edges, measured at both mentioned beamlines, 100 meV at the C *K* edge, recorded at the PM-3, and 150 meV at the *N* K edge, taken at the UE56/2-PGM1. *P*-linearly polarized light with 95% and 99% degree of polarization was used at the PM-3 and the UE56/2-PGM1, respectively. Calibration of the photon energy was carried out by means of NEXAFS measurements of gaseous $N_2$ setting the position of the first N π* resonance to 400.88 eV.[1]

**Additional XA measurements**

*C K XA spectra of graphene on Ni/W(110)*

Graphene epitaxially grown on Ni(111) has already been subject to many studies. It is known that in segregated graphitic overlayers on Ni(111) single crystals the distances between carbon atoms are slightly expanded by about 2% compared to the ones within a graphite single-crystal plane.[2] Also the influence of defects within the graphene on the electronic properties of the system has been investigated.[3] Here, we show in Figure S1 XA spectra at the C *K* edge of one monolayer epitaxial graphene on top of 30 ML Ni/W(110), taken at 300 K with *p*-linearly polarized light and an angle of 90° (black line), 55° (blue line), and 20° (red line) between the incoming X-ray wave vector and the surface. The spectral feature at 285.4 eV, exhibiting a characteristic shoulder at 287.0 eV, belongs to π* transitions, whereas contributions to the spectra, lying above 289 eV, can be ascribed to σ* transitions. The flat adsorption of the graphene sheet on Ni(111), which is known from STM investigations,[4,5] determines the symmetry of the final states, involved in the absorption process. The σ and π orbitals are oriented along and perpendicular to the graphene layer, respectively. This is reflected by the pronounced angle dependence of the corresponding resonances. The spectrum taken at 20° grazing incidence is dominated by the π* resonance, whereas the one taken at normal incidence displays almost exclusively contributions of σ* resonances.[6] The presence of two chemically nonequivalent C atoms within the graphene layer explains the appearance of diverse contributions within the π* region. π orbitals of the C atoms hybridize in a different way with the Ni 3*d* states, depending on the number of nearest neighbor Ni atoms.[7]

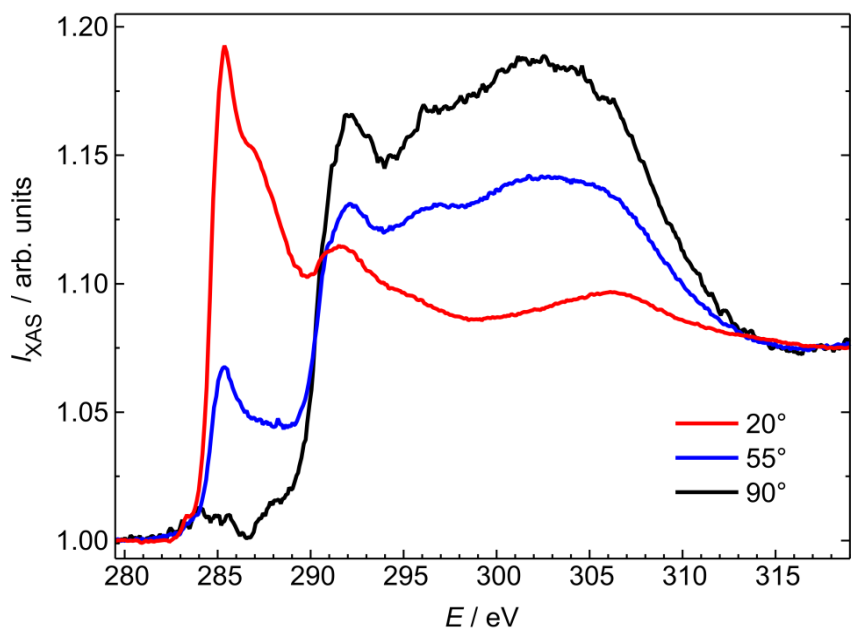

Figure S1: C *K* XA spectra of graphene on Ni/W(110) measured at 300 K with *p*-linearly polarized light and an angle of 20° (red line), 55° (blue line), and 90° (black line) between the X-ray wave vector and the surface.

*Ni $L_{2,3}$ XA and XMCD spectra of CoOEP on graphene/Ni/W(110)*

For the purpose of determining the remanent magnetization of the Ni film underneath the molecules and the graphene, XAS and XMCD spectra have been measured at the Ni $L_{2,3}$ edges. Figure S2 shows helicity-averaged Ni $L_{2,3}$ XA (a) and corresponding XMCD (b) spectra of 0.7 ML CoOEP on graphene/Ni/W(110). They have been recorded at 30 K and an angle of 20° between the X-ray wave vector and the surface. The spectra belong to the series of measurements shown in Fig. 1 of the manuscript and are taken at zero external magnetic field at remanence of the ferromagnetic film. Evaluating the spin and orbital magnetic moments by applying the sum rules [8,9] gives an effective spin magnetic moment $m_{s,eff}$ = $(0.69 \pm 0.05)$ $\mu_B$ and an orbital magnetic moment $m_l$ = $(0.07 \pm 0.01)$ $\mu_B$ for Ni. Thereby the number of holes was assumed to be 1.4,[10] and the angle of incidence (20°) as well as the degree of circular polarization (85%) have been considered. The bulk value of Ni for the spin magnetic moment (0.62 $\mu_B$) [11] is slightly lower than our experimental value for $m_{s,eff}$. This demonstrates magnetic saturation of the Ni film, as expected at these temperatures. As our presented measurements are not performed at the so called magic angle of 35.3° grazing incidence for circularly polarized light, the experimental XMCD spectrum does not present the isotropic one, as it also carries information about the anisotropy of the spin density. This is caused by the reduced symmetry at the Ni-graphene interface, and usually described by the so-called $T_z$ term. It could be a possible reason for the slightly higher than expected effective spin magnetic moment.

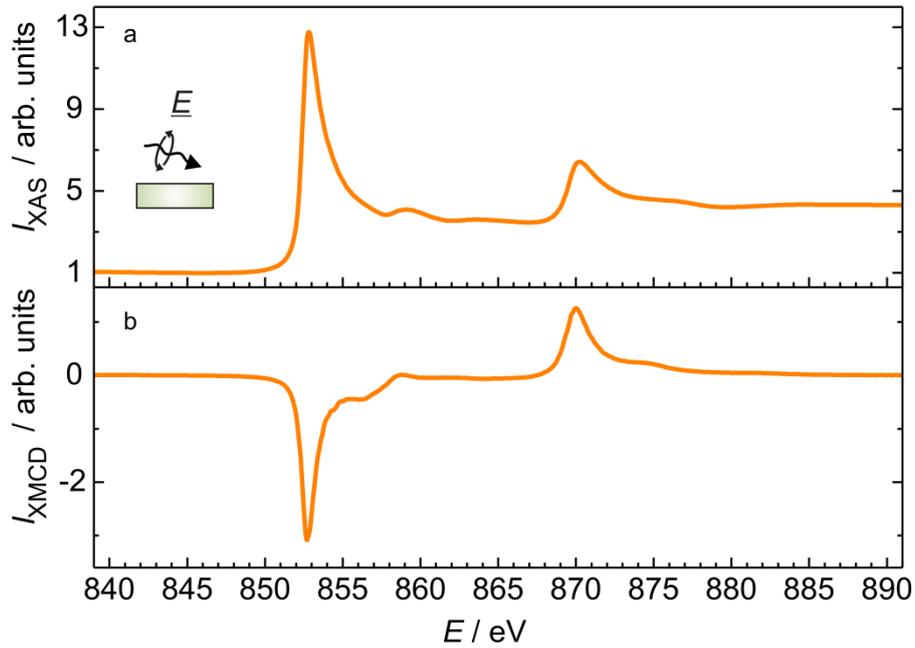

Figure S2: Helicity-averaged Ni $L_{2,3}$ XA (a) and corresponding XMCD (b) spectra of 0.7 ML CoOEP on graphene/Ni/W(110) measured at 30 K and an angle of 20° between the X-ray wave vector and the surface.

*C K and N K XA spectra of CoOEP on graphene/Ni/W(110)*

Angle-resolved N *K* edge XA spectra of 0.7 ML of CoOEP as well as angle-dependent C *K* edge XA spectra of 0.7 ML of CoOEP molecules on graphene/Ni/W(110) were measured with *p*-linearly polarized X rays, in order to gain insight into the adsorption character and geometry of CoOEP.

Figure S3 displays C *K* edge XA spectra which were recorded at angles of 25° (red line) and 90° (black line) between the incoming X-ray wave vector and the surface. In the photon energy range up to 289 eV four π* resonances are detected. The first three at photon energies of 284.2, 285.0, and 285.7 eV are only separately resolved in the spectrum taken at grazing incidence (red line). The fourth at an energy of 287.8 eV is more pronounced at grazing incidence. Above 289 eV, σ* resonances contribute to the spectra. Up to 296.8 eV, the absorption intensity is higher for grazing incidence with a maximum at 292.4 eV, while between 296.8 eV and 306.1 eV it is higher for normal incidence. These C *K* edge XA spectra of CoOEP on graphene/Ni/W(110) and theoretical C *K* edge XA spectra of a free CoOEP molecule, calculated by StoBe cluster calculations,[12] have a close overall similarity. Following this comparison and assuming a weak interaction between the CoOEP molecules and the substrate, as shown by our DFT+*U* calculations, it is possible to assign the π* resonances to excitations from C 1*s* levels, originating from carbon atoms inside the porphyrin macrocycle. In contrast, the σ* resonance with a maximum at 292.4 eV is mainly caused by excitations from C 1*s* levels arising from carbon atoms within the ethyl groups, which are directly bonded to the macrocycle. The angular dependence of the π* resonances, still visible for measurements at normal incidence, matches to a near-out-of-plane orientation of the C π* orbitals. Furthermore, the angular dependence of the σ* resonance with a maximum at 292.4 eV is an indication that the ethyl end groups are not aligned in the molecular plane.

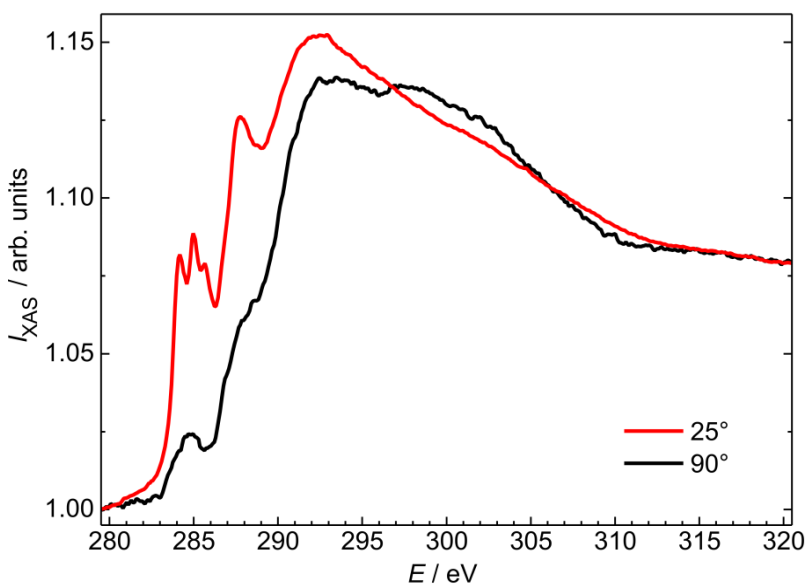

Figure S3: C *K* XA spectra of 0.7 ML CoOEP on graphene/Ni/W(110) measured at 300 K with *p*-linearly polarized light and an angle of 25° (red line) and 90° (black line) between the X-ray wave vector and the surface.

Angle-dependent N *K* edge XA measurements, presented in Fig. S4, were recorded at angles of 20° (red line), 55° (blue line), and 90° (black line) between the surface and the incoming X-ray wave vector. The spectra are built up by two π* resonances with maxima at 399.2 eV and 402.2 eV, being more pronounced at 20° grazing incidence, and a broad σ* resonance with a maximum at 407.0 eV, which is more dominant at normal incidence. Due to such an angular dependence, it can be concluded that the antibonding π* orbitals are almost perpendicularly oriented to the surface, while the antibonding σ* orbitals lie nearly parallel to the substrate. This as well as the above mentioned angular dependence at the C *K* edge fits to a flat adsorption of CoOEP, while intensity in the π* region for measurements under normal incidence can be explained by a slightly buckled shape of the porphyrin macrocycle. The N *K* edge XA spectrum taken at the so-called magic angle of incidence for linearly polarized X rays (55°, blue line), for which the dependence on the orientations of the molecular orbitals cancels out, closely resembles the N *K* XA spectrum of a polycrystalline bulk sample.[13] This again supports the assumption of a weak interaction between the CoOEP molecules and the substrate.

The DFT+*U* calculations, presented in the main article, are consistent with these results. They reveal that the molecules adsorb flat on the surface, while the porphyrin macrocycles are slightly distorted due to the interaction with the substrate (see Fig. 3 of the main article).

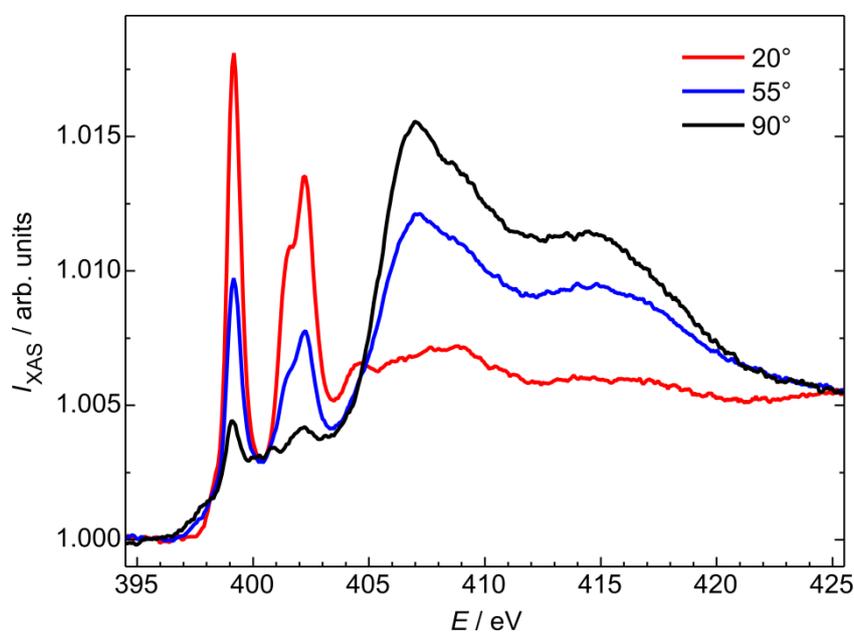

Figure S4: N *K* XA spectra of 0.7 ML CoOEP on graphene/Ni/W(110) measured at 300 K with *p*-linearly polarized light and an angle of 20° (red line), 55° (blue line), and 90° (black line) between the X-ray wave vector and the surface.

*Co $L_3$ XA spectra of CoOEP of a polycrystalline bulk sample, on graphene/Ni/W(110), and on Ni/Cu(100)*

Figure S5 shows isotropic Co $L_3$ XA spectra, recorded with *p*-linearly polarized light, of a CoOEP polycrystalline bulk sample (green line) and of 0.6 ML and 0.7 ML of CoOEP molecules adsorbed on graphene/Ni/W(110) (blue line) and Ni/Cu(100) (brown line), respectively. The bulk sample is measured at normal incidence, while the other two are taken at 55° grazing incidence. The main maxima of each spectrum are scaled on each other and the spectra are vertically shifted for clarity. The fine structure at the Co $L_3$ edge is built up by three peaks at 777.5 eV, 779.2 eV, and 780.4 eV in the case of the bulk sample as well as for CoOEP on graphene/Ni/W(110). From this line shape a +2 valency and a $d^7$ low-spin state of the Co ions is deduced for both cases by comparison to literature,[14] in agreement with the results of our DFT+$U$ calculations. The close resemblance of the two spectra points towards a very weak interaction of Co 3$d$ electronic states and the graphene-covered Ni films, which is also in line with our DFT+$U$ calculations. Nevertheless, an antiferromagnetic coupling between the Ni substrate and the Co ions is established, as demonstrated in the manuscript.

In contrast, CoOEP molecules adsorbed on a bare Ni film reveal a strongly modified Co $L_3$ XA spectrum compared to the one of the bulk sample. The fine structure at the Co $L_3$ edge now exhibits a main peak at 779.7 eV and a shoulder at 777.6 eV. The contraction of the multiplet structure, in comparison with the polycrystalline bulk sample, clearly discloses a modification of the Co electronic structure upon the adsorption of the molecules on the bare Ni film.[15] Co states hybridize with electronic states of the reactive ferromagnetic substrate, whereas on the graphene-protected Ni films no covalent bonds between the adsorbate and the substrate are established.

The extremely close similarity of the Co $L_3$ XA spectra of CoOEP bulk material and CoOEP on graphene/Ni together with the comparatively strongly modified spectrum for CoOEP on Ni/Cu(100), further reveals that a vast majority of porphyrin molecules, if not all, are adsorbed on the graphene-protected Ni substrate. The line-shape of the Co $L_{2,3}$ XA and XMCD spectra for CoOEP on graphene/Ni(111) are exactly reproducible for many different sample preparations. This verifies the successful formation of a complete graphene layer before deposition of the molecules. If part of the molecules would be adsorbed on an incompletely formed graphene layer, experimental results would vary between different preparations since the amount of imperfections would hardly be reproduced.

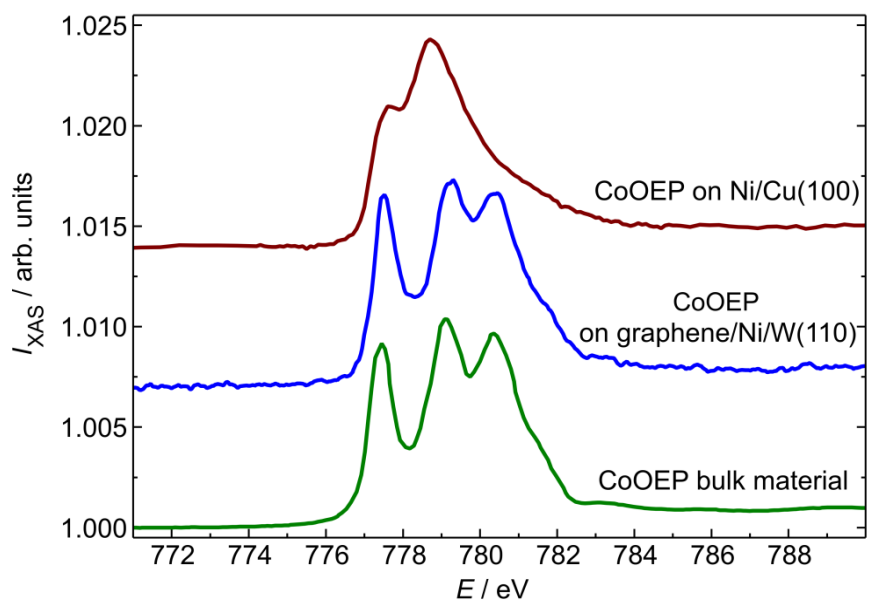

Figure S5: Co $L_3$ XA spectra of CoOEP bulk material (green line), of 0.7 ML CoOEP on graphene/Ni/W(110) (blue line), both measured at 300 K, and of 0.6 ML CoOEP on Ni/Cu(100) (brown line), recorded at 100 K. The spectra are taken with $p$-linearly polarized light at normal incidence for the bulk material, and at 55° grazing incidence for CoOEP on graphene/Ni/W(110) as well as on Ni/Cu(100). The main maxima of each spectrum are scaled to each other for comparison. The spectra are vertically offset for clarity.

**Results of density-functional theory +*U* calculations**

*Adsorption geometries of graphene on Ni(111)*

Six potential adsorption geometries for graphene on Ni(111) have been identified previously. [16,17] These geometries are denoted as the bridge-top, top-fcc, top-hcp, hcp-fcc, bridge-hcp and bridge-fcc geometry, respectively. In the top-fcc geometry carbon atoms from the two sublattices of graphene are on top of a finishing-layer Ni atom and a third layer Ni atom, respectively, whereas in top-hcp geometry they are on top of a Ni atom from the finishing layer and one in the second layer from the top. In the bridge-top geometry a top layer Ni atom is precisely under the bridge formed by two neighboring C atoms. In the fcc-hcp structure the two C atoms are on top of Ni atoms in the $2^{nd}$ and $3^{rd}$ layers from top of the Ni(111) surface. A side view of these geometries is shown in Fig. S6. The lowest energy adsorption geometries are the bridge-top, top-fcc and top-hcp geometries. As the bridge-hcp, bridge-fcc and fcc-hcp geometries have a relatively higher total energy, we consider of the latter only the fcc-hcp geometry.

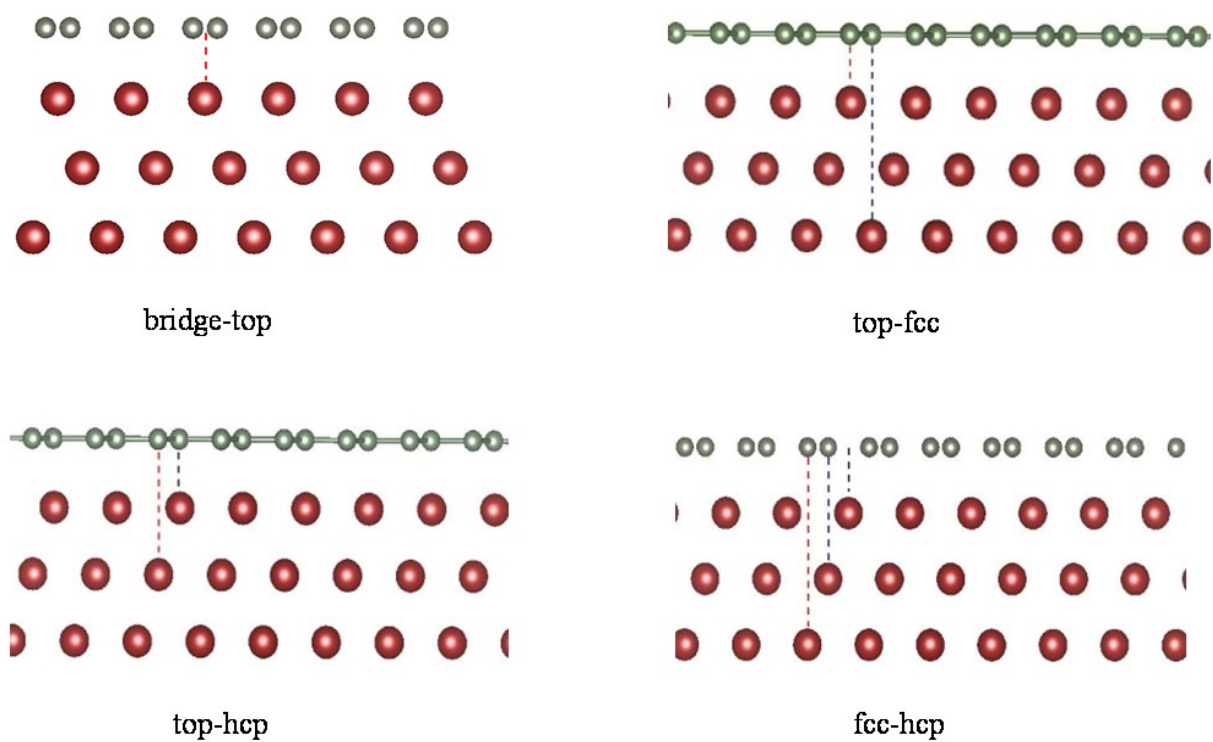

Figure S6: Illustration of the main adsorption geometries of graphene on Ni(111). From top-left to lower-right: the bridge-top, top-fcc, top-hcp and fcc-hcp geometries.

As mentioned in the main text, the adsorption of Co porphine (CoP) on each of these four graphene/Ni(111) configurations was investigated by *ab initio* DFT+*U* calculations (which included VdW corrections). Our calculations predict that CoP is adsorbed flat on the surface, and that there is only a single adsorption minimum for CoP on graphene/Ni(111) for each of these configurations. In all of them the distance of the Co central ion to the graphene layer is similar. These distances are 3.51, 3.34, 3.26, and 3.26 Å for the bridge-top, top-fcc, top-hcp, and fcc-hcp configurations, respectively. The optimized graphene – Ni top layer distances are computed as 2.08, 2.07, and 2.10, and 2.03 Å for the bridge-top, top-fcc, top-hcp, and fcc-hcp geometries, respectively. These values are in good agreement with previous calculations as well as with the experimental value of (2.11 ± 0.07) Å.[17,18,19] The computed spin of CoP is $S = 1/2$ for each of the geometries. As mentioned in the main text, a significant difference is found for the magnetic coupling of CoP to Ni in these four configurations. For the bridge-top and fcc-hcp configurations an antiparallel coupling of the Co and Ni spins is obtained, with a relatively strong coupling energy of a hundred meV for the bridge-top configuration but a weak energy of a few meV for fcc-hcp. For the top-fcc and top-hcp geometries a strong parallel coupling of the Co and Ni spin moments is predicted (with coupling energies of a hundred meV, obtained with VdW corrections). On this account, a top-fcc or top-hcp adsorption geometry of graphene on the Ni layer do not show the experimentally observed behavior. As the bridge-top configuration has the lowest total energy and provides the antiferromagnetic coupling in agreement with the experimental observations, this configuration emerges as most likely.[17]

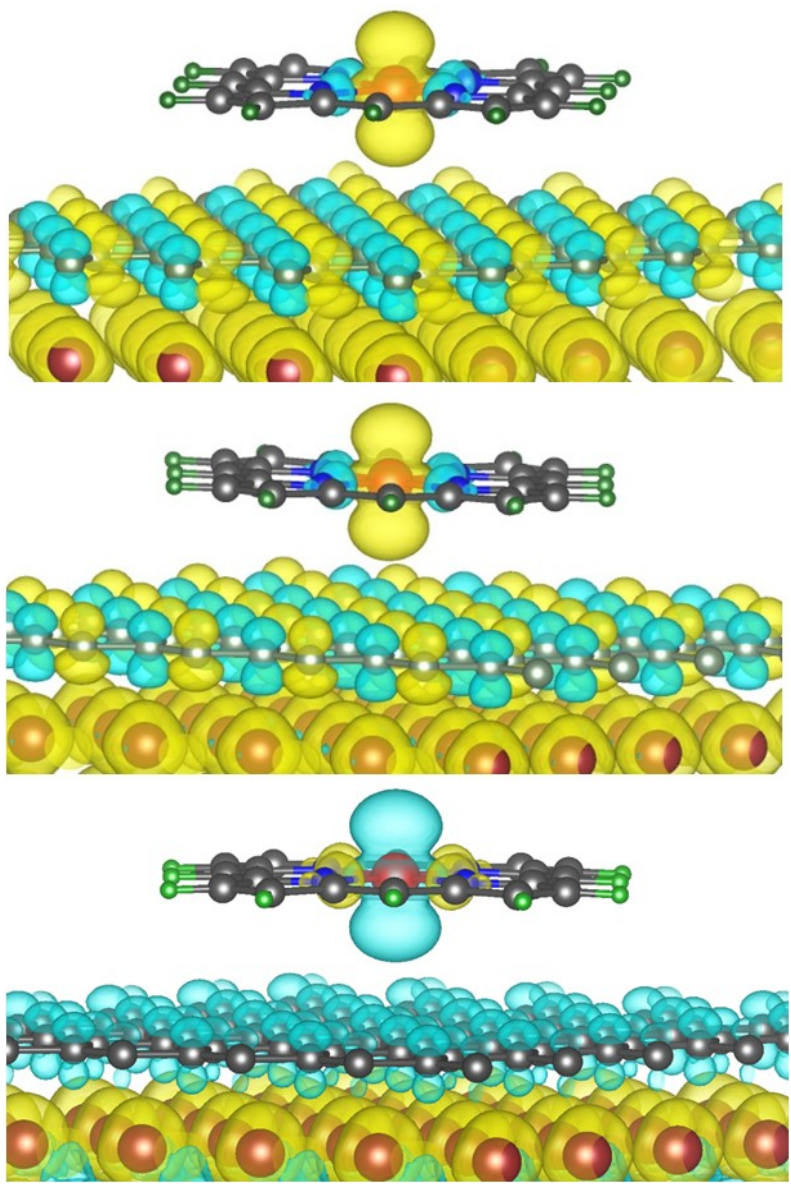

Figure S7: *Ab initio* computed magnetization densities for Co porphine on graphene/Ni(111) for different adsorption geometries of graphene. Top panel: top-fcc adsorption geometry, middle-panel: top-hcp adsorption geometry, bottom-panel: fcc-hcp adsorption geometry of graphene on Ni(111). The bright yellow hypersurfaces show contours of positive magnetization densities, the light blue hypersurface shows contours of negative magnetization density. Note the alternating positive/negative magnetization densities on the graphene C atoms in the top-fcc and top-hcp configurations and the negative magnetization density in fcc-hcp configuration. [Used isosurface values: 30 e Å$^{-3}$]

In Figure S7 we show *ab initio* computed magnetization densities for the top-fcc, top-hcp, and fcc-hcp configurations. The ferromagnetic coupling for the top-fcc and top-hcp configurations can be clearly seen from the yellow hypersurfaces around both the Co and Ni atoms. The C atoms of the two graphene sublattices acquire alternatingly a positive and a negative spin polarization, due to the interaction with the underlying Ni layer. This induced spin-polarization in graphene is markedly different from that obtained for the fcc-hcp configuration, in which all graphene C atoms acquire a small negative spin-polarization. In the top-fcc and top-hcp arrangements the magnetic coupling mediated by overlap of the porphine π orbitals and the spin-polarized graphene $p_z$ orbitals follows from the selfconsistently calculated position of the metal-organic molecule on the graphene.. The proximity of some of the C atoms to Ni top atoms induces a larger parallel spin moment on these C atoms, giving overall a net induced parallel moment on graphene. The latter induces again an antiparallel spin-polarization on the nitrogen atoms which is hence also antiparallel to that of Ni. The final chain in the exchange path, the antiparallel coupling between the spin-polarizations on the nitrogen atoms and on the Co ion, is the same irrespective of the adsorption geometry. For the fcc-hcp configuration there is also the weak antiparallel coupling between the graphene $p_z$ and porphine π orbitals, but as there is now a negative spin-polarization on the graphene, an antiparallel coupling between Co and Ni results.

*Spin-resolved density of states of the cobalt 3d orbitals*

According to crystal field theory, the 3*d* orbitals of the Co ions transform as $d_{3z^2-r^2}$, $d_{x^2-y^2}$, $d_{xy}$, and $d_{yz,xz}$ for CoOEP adsorbed on a graphene-protected Ni surface. The calculated orbital-projected spin-resolved density of states (DOS) of CoP on graphene on Ni(111) in bridge-top geometry is shown in Fig. S8. The seven 3*d* electrons of Co are distributed over the filled $d_{xy}$ and $d_{yz,xz}$ orbitals and the half-filled $d_{3z^2-r^2}$ orbital. However, a small partial hole density is present on the $d_{yz,xz}$ orbitals. The hybridization of these Co *d* orbitals can be identified from a comparison to Fig. 3 in the main text, in which the nitrogen *p* DOS is given, too. Hybridization of the $d_{3z^2-r^2}$ and $d_{x^2-y^2}$ orbitals with nitrogen *p* orbitals occurs at energies of about 5 to 6 eV below the Fermi energy ($E_F$), and a hybridization of the $d_{yz,xz}$ orbitals occurs in the energy range of 1 to 4 eV below $E_F$.

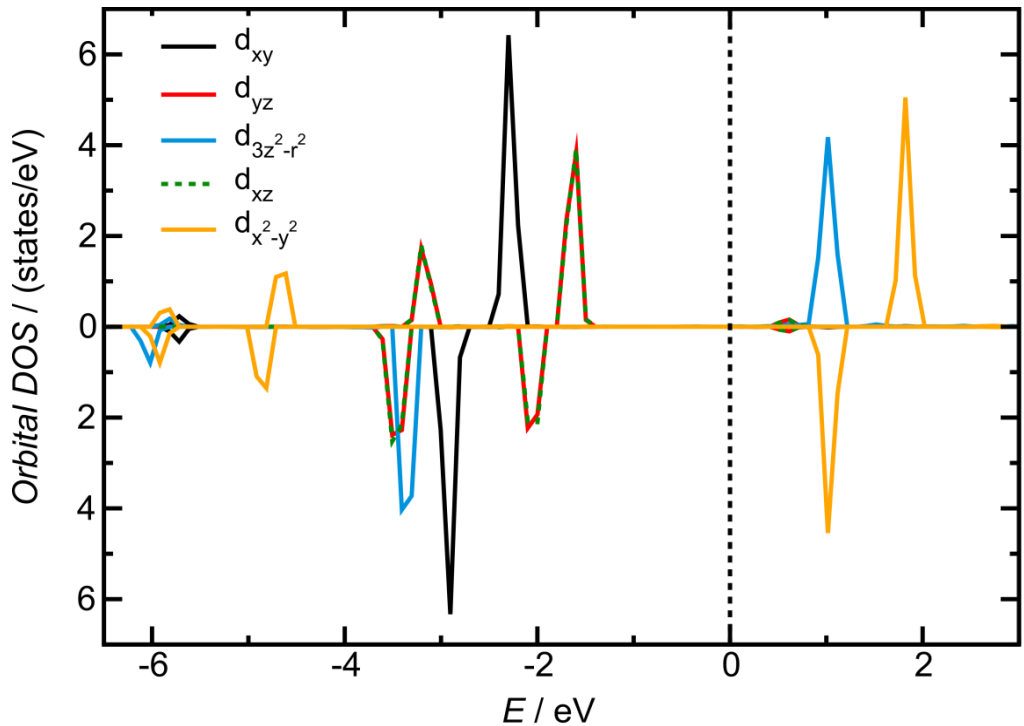

Figure S8: Spin-resolved density of states of the Co 3*d* orbitals for Co porphyrin molecules adsorbed on graphene/Ni with bridge-top configuration, as obtained from DFT+*U* calculations. Positive orbital DOS corresponds to spin up, negative atomic DOS to spin down.

**References**

[1]  R. N. S. Sodhi, V. P. Sahi, M .W. Mittelman *J. Electron Spectrosc. Relat. Phenom.* **2001**, *34*, 363.

[2]  R. Rosei, M. De Crescenzi, F. Sette, C. Quaresmi, A. Savoia, P. Perfetti, *Phys. Rev. B* **1983**, *28*, 1161.

[3]  J. Lahiri, Y. Lin, F. Sette, P. Bozkurt, I. I. Oleynik, M. Batzill, *Nat. Nanotechnology* **2010**, *5*, 326.

[4]  U. Usachov, A. M. Dobrotvorskii, A. Varykhalov, O. Rader, W. Gudat, A. M. Shikin, V. K. Adamchuk, *Vacuum* **1995**, *46*, 1101.

[5]  B. Bäcker, G. Hörz, *Phys. Rev. B* **2008**, *78*, 085403.

[6]  Weser, M. Y. Rehder, K. Horn, M. Sicot, M. Fonin, A. B. Preobrajenski, E. N. Voloshi-

na, E. Goering, Yu. S. Dedkov, *Appl. Phys. Lett.* **2010**, *96*, 012504.

[7] J. Rusz, A. B. Preobrajenski, May Ling Ng, N. A. Vinogradov, N. Mårtensson, O. Wessely, B. Sanyal, O. Eriksson, *Phys. Rev. B* **2010**, *81*, 073402.

[8] P. Carra, B.T. Thole, M. Altarelli, X. Wang, *Phys. Rev. Lett.* **1993**, *70*, 694.

[9] B.T. Thole, P. Carra, F. Sette, G. van der Laan, *Phys. Rev. Lett.* **1992**, *68*, 1943.

[10] P. Srivastava, N. Haack, H. Wende, R. Chauvistré, K. Baberschke, *Phys. Rev. B* **1997**, *56*, R4398.

[11] E. P. Wohlfahrt, in *Ferromagnetic Materials*, edited by E. P. Wohlfahrt (North-Holland, Amsterdam, 1980), Vol. 1.

[12] C. Guo, L. Sun, K. Hermann, C. F. Hermanns, M. Bernien, W. Kuch, *J. Chem. Phys.* **2012**, *137*, 194703.

[13] M. Bernien, *X-ray absorption spectroscopy of Fe complexes on surfaces*. PhD Thesis, Freie Universität Berlin (2010).

[14] P. L. Cook, X. Liu, W. Yang, F. J. Himpsel, *J. Chem. Phys.* **2009**, *131*, 194701.

[15] M. Fanetti, A. Calzolari, P. Vilmercati, C. Castellarin-Cudia, P. Borghetti, G. Di Santo, L. Floreano, A. Verdini, A. Cossaro, I. Vobornik, E. Annese, F. Bondino, S. Fabris, A. Goldoni, *J. Phys. Chem. C* **2011**, *115*, 11560.

[16] Yu. S. Dedkov, M. Fonin, *New J. Phys.* **2010**, *12*, 125004.

[17] W. Zhao, S. M. Kozlov, O. Höfert, K. Gotterbarm, M. P. A. Lorenz, F. Vines, C. Papp, A. Görling, H.-P. Steinrück, *J. Phys. Chem. Lett.* **2011**, *2*, 759.

[18] F. Mittendorfer, A. Garhofer, J. Redinger, J. Klimeš, J. Harl, G. Kresse, *Phys. Rev. B* **2011**, *84*, 201401(R).

[19] Y. Gamo, A. Nagashima, M. Wakabayashi, M. Terai, C. Oshima, *Surf. Sci.* **1997**, *374*, 61.